\documentclass[a4paper]{llncs}
\usepackage{etex}
\usepackage{amsmath, verbatim, amssymb, graphics, tabularx,  amsfonts, fixltx2e,
  color, rotate, listings, float, theorem, curves, cite, latexsym,  hyperref, fancyvrb, alltt, todonotes}
\usepackage{subcaption}
\usepackage{url, breakurl, paralist}
\usepackage{graphicx, multirow, multicol, framed, longtable, array} 
\usepackage[outdir=./]{epstopdf}
\usepackage[ruled, lined, boxed, longend, linesnumbered]{algorithm2e}
\usepackage[multiple]{footmisc}
\usepackage[htt]{hyphenat}

\newcommand\footnoteurl[1]{\footnote{\scriptsize\url{#1}}}

\usepackage{booktabs}
\begin{document}
\mainmatter
\title{Wikiwhere: An Interactive Tool for Studying the Geographical Provenance of Wikipedia References}

\author{Martin K\"{o}rner\inst{1} \and
  Tatiana Sennikova\inst{1} \and Florian Windh\"{a}user \inst{1} \and Claudia Wagner\inst{1,2} \and Fabian Fl\"{o}ck \inst{2}}
  \authorrunning{Martin K\"{o}rner et al.}
 \institute{University of Koblenz-Landau, Germany \and GESIS - Leibniz Institute for the Social Sciences, Germany
}
\maketitle

\section{Introduction and problem statement}
\label{sec:introduction}
\begin{figure}[b!]
\makebox[\textwidth][c]{
\begin{subfigure}[b]{0.5\textwidth}
  \includegraphics[width=\textwidth]{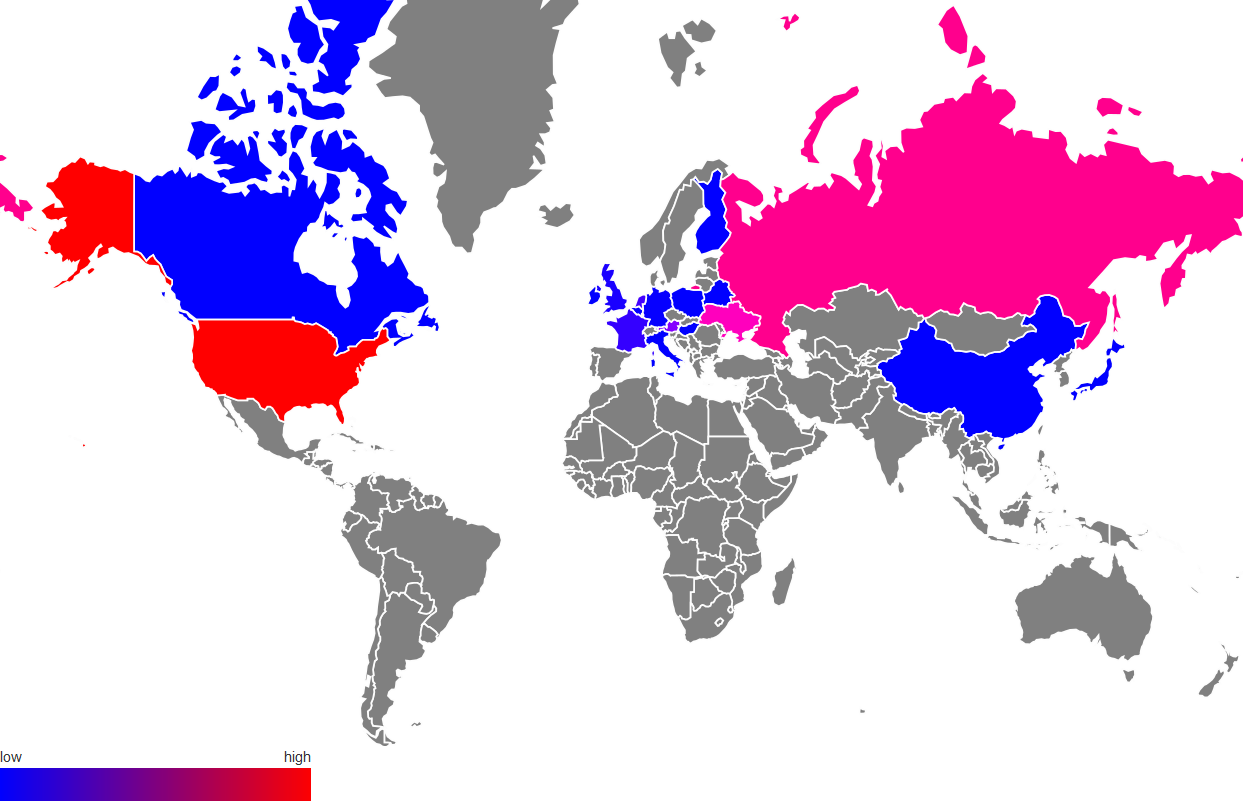}
    \caption{English version}
  \end{subfigure}
  \hfill
  \begin{subfigure}[b]{0.5\textwidth}
    \includegraphics[width=\textwidth]{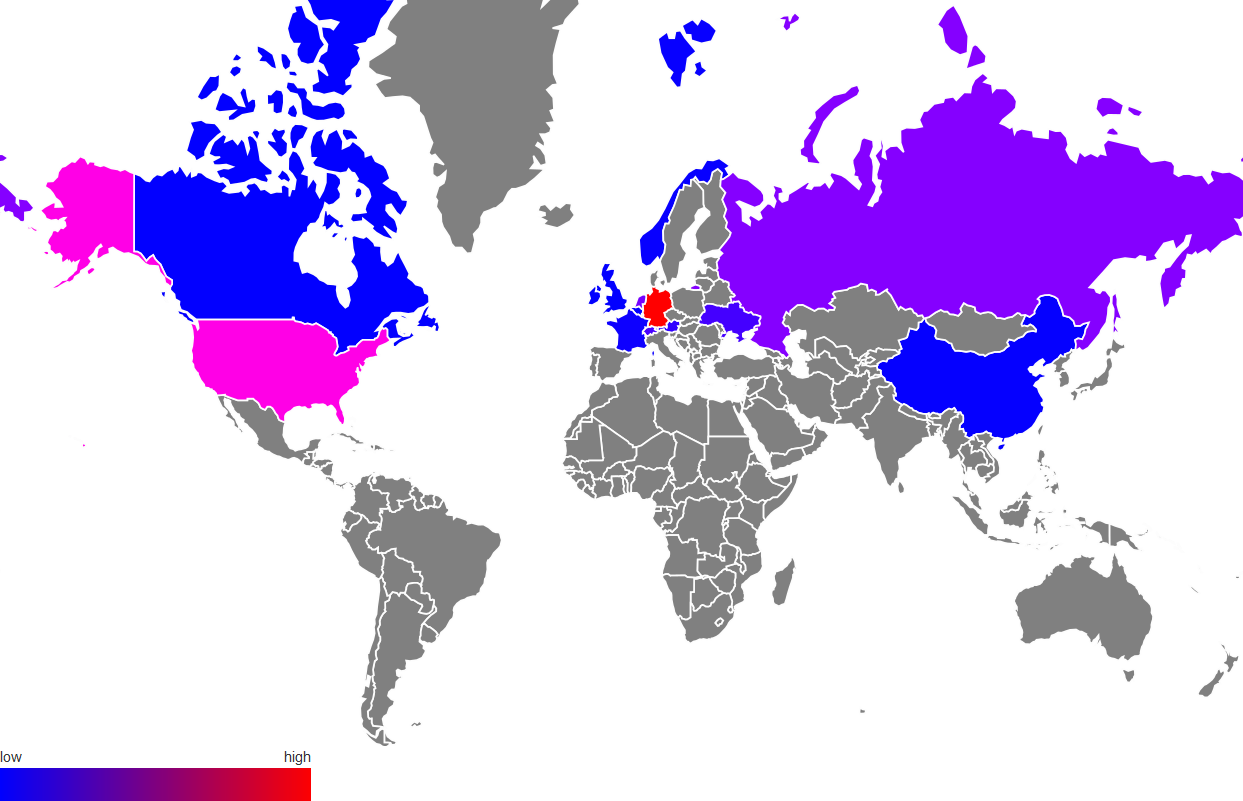}
    \caption{German version}
  \end{subfigure}
}
\caption{Comparison of the Wikiwhere heat maps of the (a) English and (b) German versions of the Wikipedia article on the "Annexation of Crimea by the Russian Federation" (English title). Blue represents least links from a country, red most, grey none. Open \url{https://goo.gl/YgJx6O} and \url{https://goo.gl/pV0Mqp} for comparison.}
\label{fig:screenshots}
\end{figure}

Wikipedia articles about the same topic in different language editions are built around different sources of information. For example, one can find very different news articles linked as references in the English Wikipedia article titled ``Annexation of Crimea by the Russian Federation'' than in its German counterpart (determined via Wikipedia's language links). Some of this difference can of course be attributed to the different language proficiencies of readers and editors in separate language editions; yet, although including English-language news sources seems to be no issue in the German edition, English references that are listed do not overlap highly with the ones in the article's English version. Remarkably, the German version, compared to its English counterpart, includes a notably higher imbalance in favor of Russian sources against Ukrainian ones, and also a lesser overall ratio of Ukrainian and Russian sources in relation to the native language of the Wikipedia edition  (cf. Figure \ref{fig:screenshots}) 
 -- although many of these pages are written in English and can be easily included in the German article.
Such patterns could be an indicator of bias towards certain national contexts when referencing facts and statements in Wikipedia. However, determining for each reference which national context it can be traced back to, 
and comparing the link distributions to each other is infeasible for  casual readers or scientists with non-technical backgrounds.

\textit{Wikiwhere} answers the question where Web references stem from by analyzing and visualizing the geographic location of external reference links that are included in a given Wikipedia article. Instead of relying solely on the IP location of a given URL
our machine learning models consider several features.

Closely related is the work by Sen et al. \cite{sen_barriers_2015} that investigates, among other aspects, the geo-provenance of URLs in Wikipedia articles describing geographic locations. A related visualization is also available.\footnote{\url{http://shilad.com/localness/index.html}}

\vspace{-0.1cm}

\section{Interface and usage}
Wikiwhere is available at \url{http://wikiwhere.west.uni-koblenz.de}.\footnote{\url{http://wikiwhere.west.uni-koblenz.de/about.php} provides additional up-to-date information.} Given a valid Wikipedia article URL of any language edition, the tool returns a classification of all references on the requested article page into countries of origin, determined by our machine learning model. The results are displayed topmost on a heat map (cf. Figure \ref{fig:screenshots}) and further down in a  bar chart. Additional bar charts show the distribution of links over countries when using only single features of our model (e.g. just IP address) and finally, all references and their location result are individually listed. By opening several language equivalents of an article, the user can thus easily compare different editions.


It is also possible to access the plotted results via URL parameters, and preprocessed analyses can be accessed via the "Articles" tab on the website. The source code is available under a free license from GitHub (\url{https://github.com/mkrnr/wikiwhere}) and can also be easily employed to classify references for research purposes beyond our visualization use case, such as statistical analyses.

\vspace{-0.2cm}

\section{Determining a reference's geographical provenance}
\vspace{-0.2cm}
We use the term \textit{reference} to refer to an URL that leads from a given Wikipedia article to another web page that is not associated with the Wikimedia foundation's  projects. 
Currently available online services aiming to determine where web sources hail from geographically often 
rely solely on IP-derived locations. But given that websites and documents are frequently hosted under arbitrary domains, in many different languages on remote servers, this might yield inaccurate results.  


To investigate this suspicion, we set up a machine learning model to infer geo-provenance.
To obtain a training set for the model, we retrieved geo-location information on Wikipedia-referenced websites from DBpedia SPARQL endpoints (\url{http://wiki.dbpedia.org/about/language-chapters}). DBpedia contains structured information that allows to link the owner of a web address to a location, or -- if such information is not explicitly encoded -- inference about the owning entity and possible parent entities (e.g., a URL of a reference belongs to a newspaper, which has no location associated, but is associated with a parent company that has location information).
In order to evaluate the accuracy of this location extraction method, we manually checked 255 locations for references that we extracted from the English DBpedia, using an explicit coding scheme. The resulting accuracy was 95\%; we thus used this data 
as our ground truth for the subsequent steps. 
Next, we randomly extracted references from Wikipedia articles and obtained their DBpedia geo-location.
For this list, which comprised a total of 233,932 URLs, we automatically retrieved the IP-location, top level domain (plus location), and website language. 
 On this data, we applied a variety of statistical learning models
. An SVM model with a one vs. one multiclass classifier consistently provided the most accurate location prediction and was selected as the eventual approach.
We trained separate prediction models for the following languages: English, German, French, Italian, Spanish, Ukrainian, Slovak, and Dutch, as for those language editions  DBpedia knowledge bases do currently exist. We also built a general model that combines the data from all DBpedia knowledge bases.
The performance of our model
 was evaluated via 10-cross fold validation. 


Table \ref{tab:accuracy} compares the accuracy of our learned model with a baseline that relies exclusively  on  IP address location. The comparison was performed on two data sets. The first includes 
``All data'', i.e., all references and their location indicated by one of the features. The second only includes references for which all three features indicate different locations and thus represents ``Difficult cases''.
As becomes apparent from Table 1, (i) using only IP location decreases location determination accuracy by 20\% in the general model (10\% to 45\% in the language-specific models) Table 2 moreover shows how much the different features contribute to the models; together, these are strong indicators that research and services should not rely on IP addresses as a sole gauge for location. (ii) This holds even more true when features differ, which is often the case nowadays when websites are hosted abroad or address an international audience in, e.g., English.

\section{Conclusion}

The main contributions of this work are: 1. An interactive tool for visual analysis of the geographical provenance of references in a Wikipedia article with tested accuracy, including source code for free reuse, and 2. the insight that IP-location-based tracking is insufficient for determining (geographical) provenance of reference documents (in Wikipedia). Further, the approach of using semantic knowledge bases as a ground truth seems to be promising for tracking other kinds of provenance of references, e.g., multi-national corporations and media networks by following links and ownership-relations between businesses. 


\setlength{\tabcolsep}{2.75pt}
\begin{longtable}[hb!]{@{}lcccccccccc@{}}
\caption{Accuracy of the learned models in comparison to a classification based on only the IP address.}\tabularnewline
\toprule
Method & General & EN & FR & DE & ES & UK & IT & NL & SV &
CS\tabularnewline
\midrule
\endfirsthead
\toprule
Method & General & EN & FR & DE & ES & UK & IT & NL & SV &
CS\tabularnewline
\midrule
\endhead
All data: Model & 0.81 & 0.81 & 0.91 & 0.90 & 0.75 & 0.96 & 0.91 & 0.96 & 0.92 & 0.98\tabularnewline
All data: IP only & 0.61 & 0.30 & 0.62 & 0.77 & 0.29 & 0.86 & 0.73 & 0.86 & 0.81 & 0.80\tabularnewline
Difficult cases: Model & 0.77 & 0.78 & 0.86 & 0.80 & 0.71 & 0.89 & 0.85 & 0.91 & 0.85 & 0.93\tabularnewline
Difficult cases: IP only  & 0.30 & 0.57 & 0.64 & 0.25 & 0.81 & 0.66 & 0.80 & 0.74 & 0.79 & 0.53\tabularnewline

\bottomrule
\label{tab:accuracy}
\end{longtable}
\setlength{\tabcolsep}{6pt}

\setlength{\tabcolsep}{4.75pt}
\begin{longtable}[b]{@{}lccc@{}}
\caption{Feature contribution over all data}\tabularnewline
\toprule
Model   & IP location & TLD location & Website Language

\tabularnewline
\midrule
\endhead
General & 61\%        & 58\%         & 25\%             \tabularnewline
EN      & 30\%        & 13\%         & 2\%              \tabularnewline
FR      & 62\%        & 73\%         & 23\%             \tabularnewline
DE      & 77\%        & 68\%         & 42\%             \tabularnewline
ES      & 29\%        & 30\%         & 7\%              \tabularnewline
UK      & 86\%        & 89\%         & 29\%             \tabularnewline
IT      & 73\%        & 70\%         & 27\%             \tabularnewline
NL      & 86\%        & 76\%         & 47\%             \tabularnewline
SV      & 81\%        & 82\%         & 29\%             \tabularnewline
CS      & 80\%        & 78\%         & 34\%  \tabularnewline
          
\bottomrule
\end{longtable}

\bibliographystyle{abbrv}
\bibliography{bibliography}

\begin{thebibliography}{1}

\bibitem{sen_barriers_2015}
S.~W. Sen, H.~Ford, D.~R. Musicant, M.~Graham, O.~S. Keyes, and B.~Hecht.
\newblock Barriers to the localness of volunteered geographic information.
\newblock In {\em Proceedings of the 33rd Annual ACM Conference on Human
  Factors in Computing Systems}, CHI '15, pages 197--206, New York, NY, USA,
  2015. ACM.

\end{thebibliography}
\end{document}